\documentclass[%
 reprint,
amsmath,amssymb,
prapplied,
]{revtex4-2}

\usepackage{mathtools}
\usepackage{graphicx}
\usepackage{dcolumn}
\usepackage{bm}
\usepackage{xcolor}
\usepackage{ulem}
\newcommand{\john}[1]{\textcolor{black}{#1}} 
\newcommand{\jon}[1]{\textcolor{black}{#1}} 
\newcommand{\richard}[1]{\textcolor{black}{#1}} 
\usepackage{hyperref}

\DeclareUnicodeCharacter{2009}{\,} 

\begin{document}

\preprint{APS/123-QED}

\title{Object detection and rangefinding with quantum states using simple detection}

\author{Richard J. Murchie, Jonathan D. Pritchard and John Jeffers}
 \affiliation{Department of Physics and SUPA, University of Strathclyde, John Anderson Building, 107 Rottenrow, Glasgow G4 0NG, United Kingdom}

\date{\today}

\begin{abstract}
In a noisy environment with weak single levels, quantum illumination can outperform classical illumination in determining the presence and range of a target object even in the limit of sub-optimal measurements based on \jon{non-simultaneous}, phase-insensitive coincidence counts. Motivated by realistic experimental protocols, we present a theoretical framework for analysing coincident multi-shot data with simple detectors. This approach allows for the often-overlooked non-coincidence data to be included, as well as providing a calibration-free threshold for inferring an object’s presence and range, enabling a fair comparison between different detection regimes. Our results quantify the advantage of quantum over classical illumination when performing target discrimination in a noisy thermal environment, including estimating the number of shots required to detect a target with a given confidence level.
\end{abstract}

\maketitle

\section{\label{sec:level1}Introduction} 
\richard{Quantum illumination (QI) describes the process of using a non-classical light source to perform optical rangefinding and object detection. The light state is often chosen to be a two-mode squeezed state generated from spontaneous parametric down conversion (SPDC). It offers an intrinsic advantage over classical illumination (CI) when the emitted signal is weak but there is strong environmental noise \cite{Lloyd2008}. Ideally entanglement in the QI source can be used to obtain up to a 6~dB enhancement over CI via optimal joint measurements \cite{Tan2008}, however the requirement for phase-sensitive measurements is technically challenging \cite{Guha2009,Zhang2013,Zhang2014,Zhang2015,Barzanjeh2015,Weedbrook2016,Zhuang2017,Zhuang2017_2,Zhuang2017_3,Xiong2017,Pirandola2018,Nair2020,Karsa2020,Karsa2020_2,Yung2020,Jo2021,Chen2022,Wei2022,Sorelli2022,Borderieux2022} or the measurement may not even be known \cite{Shapiro2009,Usha2009}. Instead, it is possible to exploit not the entanglement but the strong correlations of the photon pairs generated in the weak limit of the SPDC process to obtain a quantum advantage with a simpler detection protocol. These photon pairs have several possible correlations, including photon-number, temporal and spectral. Therefore, in essence, object detection via quantum illumination entails sending a probe state of the light field (conventionally the signal) towards a possible target object and recording the light that reaches the detection system, which may include some signal reflected off the target. The target, if it is present, sits in a noise bath of classical light, which is detected by the signal detector whether or not the target is there. When the the other mode (the idler) is measured, the non-classical correlations with the signal mode can be used to enhance the sensitivity of the signal mode measurement. An object's presence, for both QI and CI, is revealed by returned signal, otherwise that light is lost to the environment and the object's absence results in a return of solely noise. QI allows us to pick out returned signal photons from this noise more easily and so provides more information per photon sent to the target.}

More generally QI can be framed as a quantum state discrimination problem, due to the binary situation of the object present (H1) or absent (H0) hypotheses \cite{Barnett2009}. It is well-known that illumination with an entangled source can improve the distinguishability between two returned quantum states, even in entanglement-breaking conditions \cite{Sacchi2005}. \richard{Furthermore, while not optimal,} analysis of systems with independent quadrature measurements on the signal and idler show that QI retains an advantage over CI, while not necessarily being better than the best possible classical source \cite{Chang2019,Barzanjeh2020,Luong2020}. Similarly, it has been shown that QI with photon-counting and second order correlation measurements retains advantage over CI \cite{Lopaeva2013,Ragy2014,Meda2017,Gregory2020,Ortolano2023,Zhao2022}. Recently, a QI-based target detection method using non-local cancellation of dispersion has been developed \cite{Blakey2022,Franson1992}. More pertinently however, QI with simple photo-counting by click detectors has also been shown to provide a quantum enhancement \cite{England2019,Yang2020,He2020,Liu2020,Yang2021,Murchie2021,Yang2022}. This enhancement is smaller because such simple detection is sub-optimal in the sense that it does not saturate the Helstrom bound and so reveals less information about the quantum state than ideal measurement would \cite{Helstrom1969}. However, QI with photon-counting using click detectors is the easiest to implement experimentally, which suggests that it is suitable as an approach for developing a practical quantum-enhanced LIDAR.

This paper presents a model for object detection and rangefinding with quantum states using simple detection with Geiger-mode click photodetectors. While rangefinding in a QI-based detection scheme has been demonstrated before \cite{Liu2019,Frick2020,Zhuang2021,Kuniyil2022}, the method that we present treats detector data differently in that multiple detector data channels are condensed into \jon{a single} metric. \jon{This approach means the often-overlooked information from non-coincidence events can also be included to enhance state discrimination}. The method facilitates comparison between different detection schemes, for example CI and QI, for inference of an object's presence and range via \jon{a} metric whose interpretation depends on the likelihood of an object's presence. \richard{Our results quantify the advantage of quantum over classical illumination when performing target discrimination in a noisy thermal environment. We provide an operator-friendly approach to quantifying system performance via estimation of the time required to detect a target with a given confidence level.} Our experimentally-motivated theoretical framework has been applied to demonstrate the jamming resilience of quantum rangefinding \cite{mateusz23}.

The paper is organised as follows. In Sec.~\ref{sect2} an overview of the \jon{model system for CI and QI is given}. Section~\ref{sect3} describes object detection without timing information, detailing quantum hypothesis testing and the log-likelihood value (LLV) framework for interpreting detector data. In Sec.~\ref{sect4} we explore the effect of target distance on the object detection protocol. Section~\ref{sect5} describes the rangefinding protocol, incorporating expected delay and click stream matching. Finally, section~\ref{sect6} provides a discussion and outlook.
 
\section{System overview}
\label{sect2}
\subsection{Schematic of classical and quantum illumination}
\label{ch2.1}
\begin{center}\begin{figure}
\includegraphics{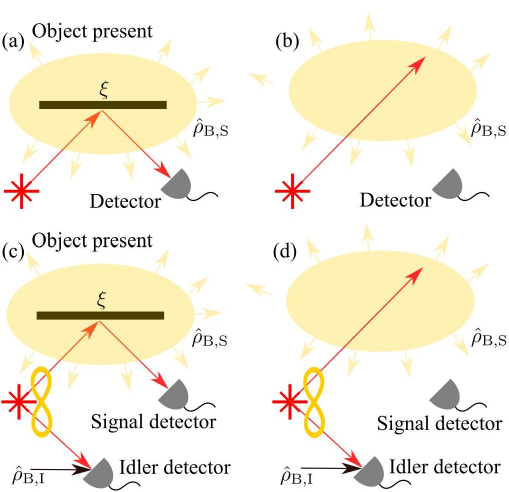}
\caption{Schematic of optical LIDAR for a target of finite reflectivity $\xi$ in a thermal background $\hat{\rho}_\mathrm{B,S}$. In the CI regime with target (a) present or (b) absent, a thermal signal beam is used to interrogate the target with a single signal detector used to measure the return field mode. In the QI regime with target (c) present or (d) absent, a photon pair-source is used to illuminate the target, with an additional detector used to directly measure the idler mode (accounting for idler background noise $\hat{\rho}_\mathrm{B,I}$) providing a coincident detection channel.}\label{schematic}
\end{figure}
\end{center}

An overview of the system for performing optical detection of a target object immersed in a thermal background using simple detection is shown schematically in Fig.~\ref{schematic} for both classical and quantum illumination \jon{regimes}. For the CI case, a thermal beam is used to interrogate the target and data provided by recording click counts on the signal detector. \john{ We consider a QI source produced by a pulsed pump laser with repetition rate $f_\mathrm{rep}$. Each pump pulse produces, via parametric downconversion \cite{Caves1985,Barnett1987}, a QI source state located centrally within a single pulse temporal window of duration $1/f_\mathrm{rep}$. The mechanism for state production is a close-to simultaneous photon pair production from one pump photon. The quasi-simultaneous nature of the pair production provides effective short term temporal correlations that are exploited to enable target detection and rangefinding via coincident detection. Outside this very short timescale the two beams are effectively uncorrelated.
Each state is described mathematically by the two-mode squeezed vacuum state (TMSV), and we assume that the pump is of a strength that it produces a TMSV with a mean photon number much smaller than one.} This state is distributed over two spatially separated modes: the signal beam and the idler beam. \john{ Within the short correlation timescale the TMSV has non-classical photon-number interbeam correlations; outside it the two beams are uncorrelated. Note that for QI we do not necessarily require a pulsed pump. The intrinsic correlations of the SPDC source enable this same framework to be applied to a continuous wave (CW) pump with the corresponding time window of a single shot equal to a coincidence detection window duration $\tau_{c}$. Note that there are other nonclassical correlations in the SPDC output that could be exploited, such as spectral correlations \cite{Frick2020}, or polarization, but we do not use these in this work. }

For \jon{CI} we assume that the signal is a \jon{pulsed thermal source} with repetition rate $f_\mathrm{rep}$, but filtered to have the the same frequency and mean photon number as the signal. We use its statistics to derive the single-shot click probabilities associated with a single pulse window duration $1/f_\mathrm{rep}$. Such a modulated temporal reference is essential for performing rangefinding in the classical system.

The detectors assumed by our model are Geiger-mode avalanche photo-diodes, insensitive to phase and which register a click or a no-click event for each experimental shot of the system, hence the term `simple detection' \cite{Renker2006}. As the signal is produced near the single-photon level the detectors are thresholded such that they can be triggered by single-photon events. This makes them appropriate for use in realistic low signal strength sensing environments.

\begin{center}\begin{figure}
\includegraphics{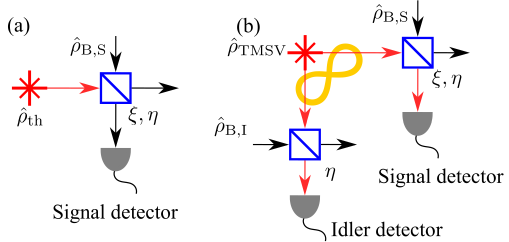}

\caption{Simplified detection model using beamsplitter transformations. (a) In the CI regime a beam splitter is used to mix the thermal signal beam $\rho_\mathrm{th}$  with the thermal background $\hat{\rho}_\mathrm{B,S}$ using parameters  $\xi$ and $\eta$ to account for finite target reflectivity when object is present and detector efficiency (b) For QI a two-mode squeezed state $\hat{\rho}_\mathrm{TMSV}$  is used, with a second beam splitter in the idler path to  account for idler background noise $\hat{\rho}_\mathrm{B,I}$ and detection efficiency $\eta$.\label{beamsplitter_diagram}}
\end{figure}
\end{center}

Losses incurred during the full target identification process are included in the model. Both detector quantum efficiency and signal attenuation from the process of probing a target object are modelled by beamsplitter transformations \cite{Barnett1998,Rohde2006}. The two input states of the beamsplitter are the background noise (or, for a detector, the source of dark counts) and the probe mode signal state. Therefore, the two beamsplitter output states are the reflected mode destined for the detector and a traced-out mode which is discarded. This partially-traced beamsplitter transformation facilitates the mixing of signal and noise, while also modelling signal loss. The mathematical beamsplitter models for QI and CI are illustrated in Fig.~\ref{beamsplitter_diagram}.

\subsection{Quantum state descriptions}
There are two easily-produced signal states typically considered for CI, the single-mode thermal state and the single-mode coherent state \cite{Mandel1995}. The density matrix for the single-mode thermal state is
\begin{equation}
    \hat{\rho}_{\text{th}}=\frac{1}{\bar{n}+1}\sum^\infty_{n=0}\left(\frac{\bar{n}}{\bar{n}+1}\right)^n\vert n \rangle\langle n \vert,
\end{equation} 
where $\bar{n}$ is the mean photon number of the state and $\vert n \rangle$ is a photon number state in the relevant mode. The thermal state has no off-diagonal density matrix elements. This is not true for the coherent state, which is characterised by a complex number $\alpha$.

If we write the coherent state in density operator form and disregard the off-diagonal elements
\begin{equation}\hat{\rho}_{\text{coh}} =  e^{-\vert \alpha \vert^2}\sum^\infty_{n=0}\frac{\vert \alpha \vert^2}{n!}\vert n \rangle\langle n \vert . \label{approx_coh}
\end{equation} 
This is valid as the detectors are insensitive to off-diagonal elements of the density matrix. This state has Poissonian photon statistics, compared with the super-poissonian photon statistics of the thermal state, which causes the coherent state to perform slightly better than the thermal state at target detection, at the expense of a little covertness. This relative reduction in covertness occurs because an intruder could measure the second order correlation function $g^2(0)$  of the coherent state, which is distinct from that of the thermal background. Furthermore, its spectral linewidth will probably be far smaller than the background noise, which is intrinsically multimode. Throughout this paper only the single mode thermal state for CI will be considered in the results and discussion. Consideration of CI as a protocol allows for performance comparison relative to QI, in order to seek out a quantum advantage. Despite the fact that the CI presented here is not at its full potential due to the exclusion of the coherent state, the differences in results presented are small in the low mean photon number regime.

The state used in the QI simple detection system is the TMSV, with density matrix 
\begin{equation}
    \hat{\rho}_{\text{TMSV}}=\frac{1}{\bar{n}+1}\sum^\infty_{n=0}\left(\frac{\bar{n}}{\bar{n}+1}\right)^n \vert n \rangle_{\text{I}}\langle n \vert\otimes\vert n \rangle_{\text{S}}\langle n \vert,
\end{equation} 
where $\text{S}$ and $\text{I}$ denote the signal and idler mode respectively. The quantum theory of generating the TMSV is typically based on using a strong pulsed coherent pump field
driving a nonlinear material. This causes spontaneous parametric down conversion, which results in the non-classical photon-number correlated state $\hat{\rho}_{\text{TMSV}}$. The mean photon number for $\hat{\rho}_{\text{TMSV}}$ depends on the second order nonlinearity of the material $\chi^{(2)}$ and the pump field mean photon number.
If only one mode of the TMSV is used and the other discarded, the state reduces to a single-mode thermal state. This facilitates covertness as an intruder who does not have access to the idler beam is ill-poised to identify this state as it has the photon-number statistics and coherence of the same form as \jon{a weak thermal background}.

The background noise of the system is modelled by a single-mode thermal state. An assumption is made that the thermal state has the same spectral properties of the signal state, otherwise it is trivial to exclude its influence on the system. The CI and the QI signal detector background noise has a mean photon number $\bar{n}_{\text{B,S}}$. The QI idler detector's background noise has a mean photon number $\bar{n}_{\text{B,I}}$.
As per Planck's law, the background noise mean photon number can be found via $\bar{n}_{\text{B}}=1/(e^{\frac{\hbar\omega}{k_{\text{B}}T}}-1)$, where $\hbar$ is the reduced Planck's constant, $\omega$ is the frequency of the mode, $k_{\text{B}}$ is Boltzmann's constant and $T$ is the temperature.

\subsection{Detector modelling and measurements}
The detectors register, for a single experimental shot, either a click or a no-click event. The probability of a click event, which is the expectation value of the product of the signal state and the click event measurement operator, or positive operator-valued measure (POVM) element \cite{Barnett2009_book}, is calculated. An increase in mean photon number incident on the detector raises the probability of a click event. The no-click POVM \john{element} is denoted $\hat{\pi}_{\times}$ and the click event POVM \john{element} $\hat{\pi}_{\checkmark}=\hat{1}-\hat{\pi}_{\times}$. Beamsplitter transformations and tracing \jon{out} of unobserved modes, as in Fig.~\ref{beamsplitter_diagram} provide the POVM elements, as derived in Appendix~\ref{beamsplitter}, where Eq.~\ref{generic_click_POVM} defines the generic click POVM element. In this simplified treatment the deleterious effects of detector dead-time, after-pulsing and timing jitter are assumed negligible due to the performance of technology available and the low mean photon numbers considered. \john{Their effects are relatively simple to include, but complicate the theory.} Then the signal detector click POVM element is
\begin{equation}
    \hat{\pi}_{\text{S}}=\hat{1}-\frac{1}{1+\bar{n}_\mathrm{B,S}}\sum^\mathrm{\infty}_{n=0}\bigl(1-\frac{\eta_\mathrm{S}\xi}{1+\bar{n}_\mathrm{B,S}}\bigr)^n\vert n \rangle \langle n \vert,
\end{equation} 
where $0\leq\xi\leq 1$ is the signal attenuation factor, which \richard{accounts} for all loss from the process of probing a target. \richard{When denoting a POVM element the $\checkmark$ subscript has been suppressed, and will be for the rest of the paper.} \richard{The system loss parameter in the signal detector channel is signified} by $\eta_\textrm{S}$, where $0\leq \eta_\textrm{S} \leq 1$, and includes includes detector quantum efficiency and coupling losses, for example. The signal detector background noise $\bar{n}_{\text{B,S}}$ is \richard{included in the POVM element} too. The inclusion of noise and inefficiencies causes the effect of click measurement to effect the projection of the state onto a mixed state, \john{ the complement of a probability-weighted thermal state} rather than onto the complement of the vacuum.

In line with the above, the CI object present click probability is
\begin{equation}
\text{Pr}_{\text{H1:CI}}=\text{Tr}\bigl(\hat{\pi}_{\text{S}} \hat{\rho}_\mathrm{th}\bigr).
\end{equation} 
On the other hand, if the target is absent, \john{the signal state at the detector is the environment state, which is strictly a thermal state. In our model this is included in the detector POVM element, so the the signal state \richard{at the detector} becomes the single-mode vacuum, as the signal is lost to the environment.} Therefore for CI, the click probability when the target is absent is
\begin{equation}
    \text{Pr}_{\text{H0:CI}}=\text{Tr}\bigl( \hat{\pi}_{\text{S}}\vert 0 \rangle\langle 0 \vert \bigr).
\end{equation} 

For QI there is another detector, the idler, which has its own POVM. \richard{The idler POVM element is}
\begin{equation}
    \hat{\pi}_{\text{I}}=\hat{1}-\frac{1}{1+\bar{n}_\mathrm{B,I}}\sum^\mathrm{\infty}_{n=0}\bigl(1-\frac{\eta_\mathrm{I}}{1+\bar{n}_\mathrm{B,I}}\bigr)^n\vert n \rangle \langle n \vert.
\end{equation} 
This POVM element includes the system loss of the \richard{idler detector channel} $\eta_\mathrm{I}$ and the measured idler detector background noise $\bar{n}_{\text{B,I}}$.

The probability of an idler click event is 
\begin{equation}
    \text{Pr}_{\text{I}}=\text{Tr}\bigl(\hat{\pi}_{\text{I}} \hat{\rho}_{\text{TMSV}} )\bigr. .
\end{equation}
 Such a click event on the idler detector conditions the signal state, which is then measured by the signal detector. In a noiseless and perfect efficiency system this conditioning heralds the presence of at least one photon in the signal mode, as measurement of an idler click conditions the signal state to have a mean photon number of $\bar{n}+1$. This conditioning also shows that the heralding gain, $1+1/\bar{n}$, is largest when $\bar{n}\ll1$\john{, such that a single signal photon is produced by an idler click}. 
 
There are two possible types of click events for the signal detector, the coincidence click and non-coincidence click. The probability of a coincidence click and non-coincidence click \john{(or alternatively the probability of a signal click given an idler click or no click)} when an object is present, are respectively 
 \begin{eqnarray}
    \text{Pr}_{\text{H1:I,1}}&=&\text{Tr}\biggl( \hat{\pi}_{\text{S}}\frac{\text{Tr}_{\text{I}}(\hat{\pi}_{\text{I}}\hat{\rho}_{\text{TMSV}})}{\text{Pr}_{\text{I}}}\biggr),\\
    \text{Pr}_{\text{H1:I,0}}&=&\text{Tr}\biggl( \hat{\pi}_{\text{S}} \frac{\text{Tr}_{\text{I}}((\hat{1}-\hat{\pi}_{\text{I}})\hat{\rho}_{\text{TMSV}})}{1-\text{Pr}_{\text{I}}}\biggr).
\end{eqnarray}
The object absent click probabilities for QI are now described. The idler click probability $\text{Pr}_{\text{I}}$ is not affected by the object and the coincidence $\text{Pr}_{\text{H1:I,1}}$ and $\text{Pr}_{\text{H1:I,0}}$ non-coincidence probabilities are now both equivalent to 
\begin{equation}
 \text{Pr}_{\text{H0}}\equiv \text{Pr}_{\text{H0:CI}}.
\end{equation}
These probabilities underpin the theory for target object detection.

\section{Object detection without timing information}
\label{sect3}
A single-shot measurement with simple click detectors can not effectively distinguish between the two cases of object present and absent. This is because when background noise is present a click in a single-shot system can either originate from the reflected signal beam or from background noise. Hence, multi-shot quantum hypothesis testing is required \cite{Wilde2013,Spedalieri2014,Wilde2017,Zhuang2020}.

Each shot of the experiment at each detector corresponds to a Bernoulli trial. A click or no click event occurs according to a click probability and generates a corresponding binomial click count distribution \cite{Papoulis1984}.
In order to infer an object's presence or absence, comparison of the relevant click probability distributions is required \cite{Garthwaite2002}. 
In the limit of many shots the binomial click distributions can be approximated as Gaussian, under the assumption \richard{that the criterion detailed in Appendix~\ref{gaussian} is satisfied}. This valid approximation greatly simplifies the analysis and computational demands and is used henceforth.

\begin{figure}[b]\begin{center}
\includegraphics{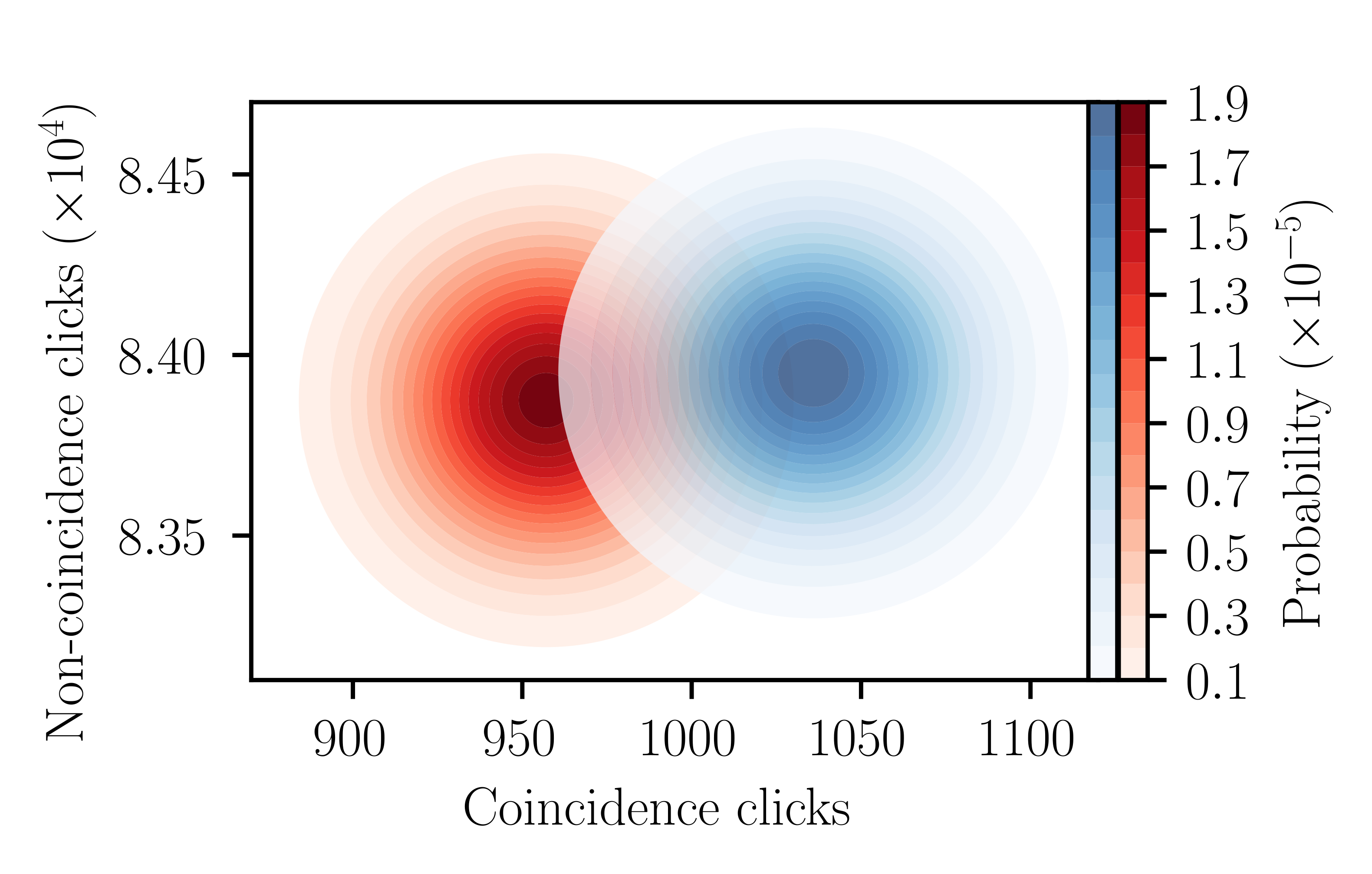} 
\caption{Quantum illumination probability distribution displaying coincidence click counts and non-coincidence click counts for both object present (coloured blue) and absent (coloured red) cases. Mean photon number of the signal state $\bar{n}=2.19\times10^{-2}$, system loss of all detectors $\eta_\mathrm{S/I}=0.5$, signal attenuation factor $\xi=8.84\times10^{-3}$,  mean photon number of background and dark counts for signal detector $\bar{n}_{\text{B,S}}=5.06\times10^{-2}$, and mean photon number of background and dark counts for idler detector $\bar{n}_{\text{B,I}}=4.49\times10^{-4}$. The number of shots is $1.76\times10^6$ and the number of idler click events displayed is $1.98\times10^4$.\label{clickprob}}\end{center}
\end{figure}

Each set of system parameters gives rise to an idler click distribution $P_{\text{I}}$, and each value in $P_{\text{I}}$ has a corresponding object present $P_{\text{H1}}$ and absent $P_{\text{H0}}$ signal click distribution. Figure~\ref{clickprob} shows the coincidence and non-coincidence click distributions for both object present and absent cases, \richard{after a set number of idler click events $k=1.98\times10^4$}.

Existing literature often overlooks non-coincidence signal clicks events as a source of useful information. In addition, the recording of all types of click is required for the rangefinding protocol discussed later in this paper. The rest of this paper will focus on QI as it is straightforward to reduce the QI theory to CI only. The protocol is explained without CI explicitly mentioned unless a performance comparison between CI and QI is made.

The make-up of a shot is a topic that has been already considered briefly in Sect.~\ref{ch2.1}. Our model assumes a perfect one-to-one mapping of pulses to shots (with a repetition rate set by $f_\mathrm{rep}$). This is perfectly fine in the low detector timing jitter regime and where count rates are slow enough that detector dead times are negligible. While it is easy to set the temporal duration of a shot for a pulsed source, as the reciprocal of the source repetition rate, it is not the case for CW systems which require $f_\mathrm{rep}$ to be artificially set \jon{through the choice of the coincidence window duration $f_\mathrm{rep}=1/\tau_c$}. However, the second order coherence function $g^2(\tau)$ can instruct what temporal duration for a shot is sensible, for both CW and pulsed. Due to the spontaneous and stimulated aspects of the nonlinear process there could be many modes within one shot temporal window size, however it suffices to model each shot to only have one mode of the source due to the nature of the detectors and the low mean photon number. As such, the click probabilities are easily found via the method presented earlier. It is clear that the mean photon numbers for the source and background are dependent on the temporal window size of a shot. Henceforth this paper focuses on a pulsed pump source, noting that extension to idler-detector-gated CW is relatively straightforward. 

\subsection{Log-likelihood value}
A click count value might infer an object's presence in one parameter regime or an object's absence in another. Hence, a framework is desired that allows for fair comparison of incoming click data between different situations.  
The log-likelihood value (LLV) forms the basis of this framework. The LLV is also appropriate for use in dealing with multi-channel detector data, as it reduces multiple channels of data pertaining to two simple hypotheses into a single value. This value also provides a simple test, in this context commonly known as the likelihood ratio test. The use of the LLV for hypothesis testing is justified by the Neyman-Pearson lemma, which states that it provides the most powerful test for a set statistical significance level \cite{Neyman1933}.

The object present $P_{\text{H1}}$ and absent $P_{\text{H0}}$ click distributions in their binomial form after $N$ shots and $k$ idler click events define the LLV after $k$ idler clicks which converts click data $\underline{x}$ into a LLV
\begin{equation}
    \Lambda(\underline{x},k)=\text{ln}\left(\frac{P_{\text{H1}}(\underline{x},k)}{P_{\text{H0}}(\underline{x},k)}\right),\label{LLV1}
\end{equation}
where $\underline{x}=(x,y)$ with $x$ the coincidence click count and $y$ the non-coincidence click count. It can be seen from the definition that $\Lambda(\underline{x},k)>0$ means that an object's presence is more likely, that $\Lambda(\underline{x},k)=0$ means that both regimes are equally as likely, and $\Lambda(\underline{x},k)<0$ infers that an object's absence is more likely. An advantage of using the LLV as a test is that it can be self-calibrating \richard{to a LLV detection threshold $d_{\text{LLV}}=0$}, as the detection decision will automatically be set according to the likelihood of an object's presence. We will consider later \jon{in Sec.~\ref{sect3}C} the effect of setting LLV decision levels on false alarm probabilities and their extension to receiver operator curves.

As the click probabilities, click data, and number of shots are all real and positive Eq.~\ref{LLV1} can be stated as a linear equation
\begin{align}
    \Lambda(\underline{x},k)&=(M_1 x+kC_1)+(M_2 y+(N-k)C_2). \label{QI_lintrans}
\end{align}
In Appendix~\ref{LLVtrans_appendix} this transformation is formulated in linear equation form and constants $M_1,\; M_2,\;C_1,\; C_2$ are defined.  
 
\begin{figure}[b]\begin{center}
\includegraphics{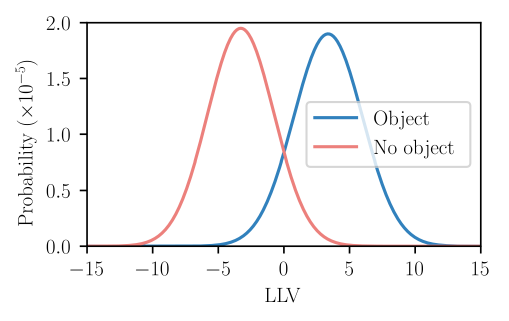}
\caption{Quantum illumination LLV probability distribution for both object present (coloured blue) and absent (coloured red) cases. $\bar{n}=2.19\times10^{-2}$, $\eta_\mathrm{S/I}=0.5$, $\xi=8.84\times10^{-3}$, $\bar{n}_{\text{B,S}}=5.06\times10^{-2}$, $\bar{n}_{\text{B,I}}=4.49\times10^{-4}$. The number of shots is $1.76\times10^6$ and the number of idler click events displayed is $1.98\times10^4$.\label{LLVdists}}\end{center}
\end{figure}

\richard{We express the statistical moments of the LLV distributions when there has been a mean number of idler clicks $k=\mu_{\text{I}}=N\text{Pr}_{\text{I}}$ in the following analysis. The object present $P_\mathrm{H1:\Lambda(x,\mu_{\text{I}})}$ and absent $P_\mathrm{H0:\Lambda(x,\mu_{\text{I}})}$ LLV distributions for idler clicks $k=\mu_{\text{I}}$ \jon{are} shown in Fig.~\ref{LLVdists}}. Assuming that the click distributions are well approximated by a Gaussian, all of the LLV distributions are Gaussian too, as linear transformations and combinations preserve normality \cite{Ghosh1969}. The results below show the object present case, but the analysis is similar for object absent. The mean and standard deviation for the object present LLV distribution after mean idler clicks as derived in Appendix~\ref{click_to_LLV} is 
 \begin{eqnarray}
    \mu_{H1:\Lambda(x,\mu_{\text{I}})}&=&N\bigl(\text{Pr}_{\text{I}}(M_1 \text{Pr}_{\text{H1:I,1}}+C_1-M_2 \text{Pr}_{\text{H1:I,0}}-C_2) \nonumber \\ &+&M_2\text{Pr}_{\text{H1:I,0}}+C_2\bigr), \\ \sigma_{H1:\Lambda(x,\mu_{\text{I}})}&=& \bigl( N(\text{Pr}_{\text{I}}(M_1^2)\text{Pr}_{\text{H1:I,1}}(1-\text{Pr}_{\text{H1:I,1}})\nonumber \\&-&M_2^2\text{Pr}_{\text{H1:I,0}}(1-\text{Pr}_{\text{H1:I,0}})) \nonumber \\ &+&M_2^2\text{Pr}_{\text{H1:I,0}}(1-\text{Pr}_{\text{H1:I,0}})\bigr)^{0.5}.
\end{eqnarray}

\subsection{Distinguishability measure}
\label{distinguish_sect}
The distinguishability measure used here describes how much overlap there is between the object present and absent LLV distributions - effectively how much confidence can be ascribed to a decision of object present or absent given an LLV. Therefore, the distinguishability measure is a figure of merit (FOM) which characterises system performance in confident detection decision-making. The use of this FOM is in contrast with other FOM's such as the signal to noise ratio (SNR) and the Cramér-Rao lower bound (CRLB) for signal loss estimation which are not directly based on decision-making. A qualitative comparison of FOMs is expanded upon in in Appendix~\ref{appendix_FOM}. \richard{The Q-function for a Gaussian distribution with LLV detection threshold $d_{\text{LLV}}$, mean $\mu$ and standard deviation $\sigma$ is \begin{equation}
    \text{Q}(d_{\text{LLV}},\mu,\sigma)=\frac{1}{\sqrt{2\pi}\sigma}\int^\infty_{d_{\text{LLV}}} e^{-\left(\frac{z-\mu}{\sqrt{2}\sigma}\right)^2}dz.
\end{equation}
We define the probability of detection and probability of false alarm respectively as \begin{equation}
\text{P}_{\text{D}}(d_{\text{LLV}})=\text{Q}(d_{\text{LLV}},\mu_{\text{H1}:\Lambda(x,k)},\sigma_{\text{H1}:\Lambda(x,k)})\label{PD}\end{equation} and
\begin{equation}
\text{P}_{\text{FA}}(d_{\text{LLV}})=\text{Q}(d_{\text{LLV}},\mu_{\text{H0}:\Lambda(x,k)},\sigma_{\text{H0}:\Lambda(x,k)}).\label{PFA}
\end{equation} The definitions in Eq.~\ref{PD} and Eq.~\ref{PFA} lead onto the distinguishability
\begin{equation}\label{single-shot_distinguishability}
    \phi=1-\left[ \left(1-\text{P}_{\text{D}}(0)\right)+\text{P}_{\text{FA}}(0)\right].
\end{equation}}

A threshold distinguishability $\phi_{t}$ is set to ensure that the effectiveness of the LLV test is consistent in different regimes. The threshold distinguishability uses the LLV distributions \richard{when there has been a mean number of idler clicks}. An assumption is made that the underlying components of the distingiushability $\phi$, $\text{P}_{\text{D}}$ and $\text{P}_{\text{FA}}$, are also consistent between regimes. \richard{Moreover, for a single regime each object present and absent LLV distribution distinguishability differs slightly according to the number of idler clicks $k$, hence the effectiveness of the LLV test differs with the number of idler clicks recorded. We ignore this discrepancy if it does not exceed the bound placed in Appendix~\ref{appendix_LLV_k_consistency}, and consider all likely LLV distributions for a regime to have a distinguishability set by the one calculated for \jon{the  mean number of} idler clicks.} Clearly, the higher $\phi_{\text{t}}$ is the more confident the detection decision. In this paper, threshold distinguishability is set to be $\phi_{\text{t}}=0.8$ in line with convention \cite{Murchie2021}.  

\richard{We define the number of shots required for threshold distinguishability}
\begin{equation}
    N_{\text{t}}=\text{int}\left(\frac{\text{F}^{-1}(\phi_{\text{t}})}{\text{Pr}_{\text{I}}}\right),
\end{equation}
where $\text{F}^{-1}(\phi_{\text{t}})$ is the inverse of the the function $\phi_{\text{t}}$, as derived in Appendix~\ref{solvedisting}. $N_{\text{t}}$ allows for different parameter regimes to be compared fairly, as increasing the number of shots $N$ increases $\phi$, when $\phi<1$.

While the system is running the return click rate can change. Hence, in order to analyse dynamically changing incoming data, a rolling window method is applied. We define the cumulative coincidence clicks after $z$ time-bins $T(z)$ and cumulative idler clicks $T_{\text{I}}(z)$, where $z$ time-bins is the number of shots that have been recorded. \richard{There is an initialisation stage while $z<N_{\text{t}}$. We have a LLV $R(z)$ for each time-bin $z$ defined as}
\begin{equation}
    R(z)=\Lambda\left(T(z)-T(z-N_{t}),T_{\text{I}}(z)-T_{\text{I}}(z-N_{t})\right), \label{rolling_shots}
\end{equation} 
for every time-bin $z\geq N_{\text{t}}$.

\begin{figure}
\begin{center}
\includegraphics{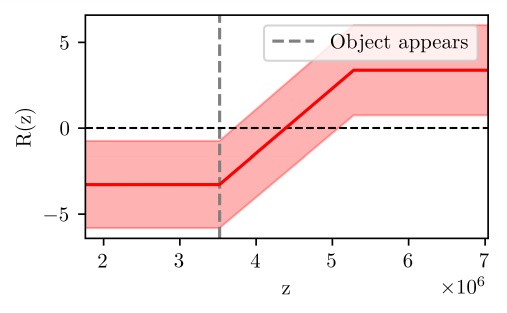}
\caption{Rolling window trajectory of QI mean LLV (solid line coloured red) with a red shaded region limited by a standard deviation of error plus and minus the mean. Regime changes from object absent to present suddenly at time-bin $z=2N_{\text{t}}$ marked by the vertical dashed line (coloured grey). $N_{\text{t}}=1.76\times10^{6}$, $\bar{n}=2.19\times10^{-2}$, $\eta_\mathrm{S/I}=0.5$, $\xi=8.84\times10^{-3}$, $\bar{n}_{\text{B,S}}=5.06\times10^{-2}$, $\bar{n}_{\text{B,I}}=4.49\times10^{-4}$.\label{rolling}}
    \end{center}\end{figure}

Figure~\ref{rolling} illustrates the change in LLV statistics between the object absent regime and the object present regime. The object suddenly appears at the time-bin denoted by the vertical dashed line. \richard{The system fully updates from object absent to present regime over $z=N_t$ time bins. Here, both regimes are separated by the threshold distinguishability $\phi_t$.}

Hence Fig.~\ref{rolling} illustrates what $\phi_{t}$ means for the distinctness of the object present and absent distributions. \richard{However, the duration of a shot is too short to practically process data in a rolling-window shot-by-shot}. Instead a rolling window based on LLV samples every $N_{\text{t}}$ shots is employed. This is elaborated upon in Sect.~\ref{rolling_sample_sect}.
\subsection{Comparing quantum illumination and classical illumination}
\begin{figure}
\begin{center}
\includegraphics{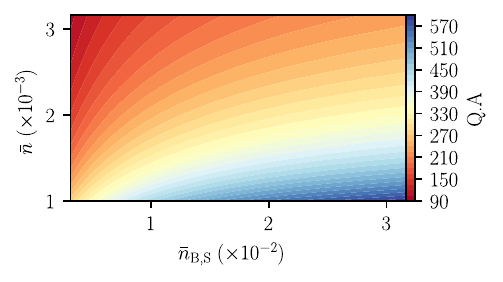} 
\caption{Contour plot of quantum advantage $\text{Q.A.}=\frac{N_{\text{t:CI}}}{{N_{\text{t:QI}}}}$ for varied background noise and signal strength. $\eta_\mathrm{S/I}=0.5$, $\xi=1.99\times10^{-2}$, and $\bar{n}_{\text{B,I}}=4.49\times10^{-4}$.\label{quantadv}}
    \end{center}\end{figure}

\richard{System performance comparison of QI and CI using click-counts directly is problematic as coincidence clicks and signal clicks are not the same, making it challenging to use click-counts as a performance metric. Our framework addresses this problem by recasting click-counts of any type into a LLV. For any system parameter regime, the object present and absent statistics for QI have a larger distinguishability than CI. We also define a quantum advantage Q.A. as the ratio of the number of shots required for CI and QI to reach threshold distinguishability. Therefore quantum advantage $\text{Q.A.}=\frac{N_{t\text{:CI}}}{N_{t\text{:QI}}}$, where $N_{t\text{:QI}}$ is $N_{t}$ shots required for threshold distinguisabitity for QI, and similarly for CI. In a physical system this quantity gives a reasonable approximation to the relative amount of time it would take for each system to determine the presence or otherwise of a target object under the same conditions.
Figure~\ref{quantadv} demonstrates that QI performs significantly better for all of the shown parameters. This advantage increases for low signal strength and high background noise regimes.}

\richard{On the other hand a common-place approach for system performance is via a receiver operator curve (R.O.C). The R.O.C can be found by varying the LLV detection threshold $d_{\text{LLV}}$ from $-\infty< d_{\text{LLV}}<\infty$ and calculating $\text{P}_{\text{D}}(d_{\text{LLV}})$ and $\text{P}_{\text{FA}}(d_{\text{LLV}})$ accordingly.}

\begin{figure}[b]\begin{center}
\includegraphics{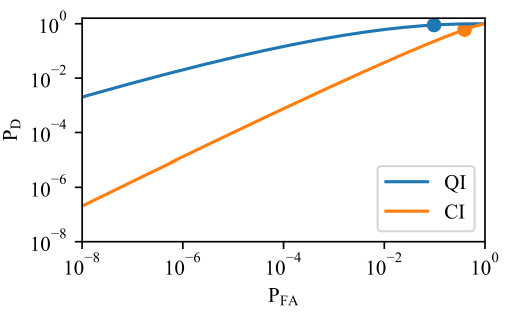}
\caption{Receiver operator curve for probability of detection $\text{P}_{\text{D}}$ and probability of false alarm $\text{P}_{\text{FA}}$ for QI and CI. The blue dot represents where the detection threshold for QI is set to LLV=0 and the orange dots represents where the detection threshold for CI is set to LLV=0. $\bar{n}=2.19\times10^{-2}$, $\eta_\mathrm{S/I}=0.5$, $\xi=8.84\times10^{-3}$, $\bar{n}_{\text{B,S}}=5.06\times 10^{-2}$, $\bar{n}_{\text{B,I}}=4.49\times10^{-4}$, and $N=1.76\times10^{6}$.\label{roc}}\end{center}
\end{figure}
\richard{Figure \ref{roc} compares two R.O.C's, one for CI and one for QI; clearly QI performs better than CI for the given parameters. To summarise, we have a method to assess system performance which is self-calibrating, and if we relax the detection threshold $d_{\text{LLV}}$ we have system performance in terms of a R.O.C.}

\section{Rangefinding without timing information}
\label{sect4}
\subsection{Signal attenuation from distance}

\richard{The process of interrogating a target object almost always results in attenuation of the signal strength}. This attenuation is dependent on inclusion of realistic effects and object properties. 

An object can be modelled as a Lambertian scatterer or a specular reflector. In this paper the Lambertian case is focused upon as the signal is scattered by the target into a large solid angle and the small solid angle subtended by the distant signal detector at the target provides a naturally-dominant attenuation factor \cite{Koppal2014}. There are other possible attenuation effects such as beam divergence and any scattering or absorption from the medium between detector and object and these can be treated similarly. If instead we were to concentrate on a cooperative specular reflector beam divergence would be the dominant attenuation that is not intrinsic to the detectors.

The signal attenuation factor $\xi$, previously introduced in this paper, is now written in terms of interaction with a Lambertian scatterer. A model for signal attenuation using the averaged Lambert's cosine law, equal hemisphere of diffusion, and the inverse square law gives
\begin{equation}
    \xi=\frac{\xi_{\text{obj}}\text{A}_{\text{d}}}{4\pi D^2},
\end{equation} 
where $\xi_{\text{obj}}$ is the intrinsic reflectivity of the object, with $\xi_{\text{obj}}=1$ as the case of a perfect reflector. The area of the detector is $\text{A}_{\text{d}}$ and distance of object from the detector is $D$. 

\subsection{Expected click count and LLV as a function of distance}
We would expect the number of shots to reach threshold $N_{\text{t}}$ to increase with detector distance $D$. \richard{The change of signal attenuation with distance requires that a specific LLV after $k$ idler clicks is defined for each each distance $D$}
\begin{equation}
    \Lambda_{D}(\underline{x},k)=\text{ln}\left(\frac{P_{\text{H1},D}(\underline{x},k)}{P_{\text{H0},D}(\underline{x},k)}\right),
\end{equation} 
\richard{with $P_{\text{H\{1,0\}},D}$ defined as the click probability distribution for an object present and absent at distance $D$ after $N_{t}$ shots, respectively. Consequently, for any distance we can process data such that the effectiveness of the LLV test remains consistent. Assuming the requirement of processing $N_t(D)$ shots for each distance $D$ is satisfied, the expected mean coincidence click count $\mu_{\text{D:I,1}}$ increases with distance due to the increased $N_{t}(D)$. The mean LLV $\Lambda_{D}(\mu_{\text{D:I,1}},\mu_{\text{I}})$ will stay approximately constant for all inspected distances, as the LLV test effectiveness is consistent.} Hence, the different distances can have their data streams directly compared, which is a valuable feature for a rangefinding protocol.

\section{Rangefinding with timing information}
\label{sect5}
\subsection{Delay and distance}
Light takes time to travel, therefore there will be a temporal mismatch between the idler and signal beams returning to the detectors due to the different path lengths travelled. In order to retrieve the light source's temporal correlations this mismatch must be accounted for, in the form of the expected delay. The incoming binary stream of click events at the detector is henceforth referred to as the idler or signal stream.  
The idler stream is set to have zero delay, as it is locally measured with a known path length. For simplicity the system we assume is monostatic. Therefore the detectors and light source are located approximately at the same location, which simplifies the relation for the expected delay. The extension to bistatic is straightforward. The expected time delay for the signal beam when an object is at distance $D$ from the detector is therefore
\begin{equation}
    t(D)=\frac{2D}{c},
\end{equation} 
where $c$ is the speed of light.

\subsection{Discretising the delay}
As the model considers the idler and signal beam data streams to be in discrete time-bins the expected delay must be discretised in order to match the signal stream with idler. The expected delay in shots for an object at distance $D$ from the detector is 
\begin{equation}
    M_{\text{delay}}(D)=\text{int}\left({t(D)}f_\mathrm{rep}\right).
\end{equation} 
Therefore, the idler stream is matched with the signal stream shifted back $M_{\text{delay}}(D)$ shots, for an inspected distance of $D$. 
The optimal spatial resolution of the system is set by the source repetition rate $f_\mathrm{rep}$ 
\begin{equation}
    D_p=\frac{c}{2 f_\mathrm{rep}}.
\end{equation}
Some error can occur due to the discretised delay not properly matching with object distances that are not integer values of the spatial resolution. This protocol's rangefinding abilities will also be tempered by the misbinning stemming from timing jitter, but this effect is neglected here, as mentioned earlier. \richard{Previous literature for simple detection based QI rangefinding describes timing jitter corresponding to $\approx10$~cm range uncertainty \cite{frickthesis}, which is smaller than the range resolution we specify.}
\richard{The temporal analogue of the optimal spatial resolution is the optimal temporal resolution $t_{\text{optimal}}(D)$ which determines how quickly our system can make a confident measurement for a distance $D$. This depends on the source repetition rate $f_\mathrm{rep}$ and the threshold distinguishability $\phi_t$ as defined in}
\begin{equation}
    t_{\text{optimal}}(D)=\frac{N_{\text{t}}(D)}{f_\mathrm{rep}}.
\end{equation}

\subsection{LLV rangefinding statistics}
{\color{blue} }
In a rangefinding scenario the distance of a possible target object is unknown, and hence the expected delay is also unknown. \richard{We systematically work through different possible distances from near to far.} 
Other than the set of model parameters which determine the expected object statistics, there are two pieces of information that the operator can control when searching for an object: how long does it take to acquire confident statistics $N_{\text{t}}$ and what delay should be used to match streams correctly $M_{\text{delay}}(D)$. Table~\ref{tab:table1} shows $N_{\text{t}}$ and $M_{\text{delay}}(D)$ \richard{for a set of inspected} distances.
\begin{table}[b]
\caption{\label{tab:table1}
Shots required $N_{\text{t}}$ for threshold distinguishability $\phi_{\text{t}}=0.8$ and return signal delay in shots for target object distance $D$ from detector $M_{\text{delay}}(D)$. Lambertian scatterer. Source pulse repetition rate $f_\mathrm{rep}=0.5$~GHz. $\bar{n}=2.19\times 10^{-2}$, $\eta_\mathrm{S/I}=0.5$, $\xi_{\text{obj}}=1$, $\bar{n}_{\text{B,S}}=5.06\times 10^{-2}$, $\bar{n}_{\text{B,I}}=4.49\times 10^{-4}$.
}
\begin{ruledtabular}
\begin{tabular}{lcdr}
\multicolumn{1}{c}{\text{Distance (m)}}&
$N_{\text{t}}$ & M_{\text{delay}}(D) & $\xi$ \\
\colrule
1.2 & 5.29$\times 10^4$ & 4 & 5.53$\times 10^{-2}$\\ 
3 & 1.76$\times 10^6$ & 10 & 8.84$\times 10^{-3}$ \\
3.3 & 2.56$\times 10^6$ & 11 & 7.31$\times 10^{-3}$ \\
6 & 2.73$\times 10^7$ & 20 & 2.21$\times 10^{-3}$ \\
CI: 3 & 3.91$\times 10^7$ & 10 & 8.84$\times 10^{-3}$ 
\end{tabular}
\end{ruledtabular}
\end{table}

\begin{figure}[b]
\begin{center}
\includegraphics{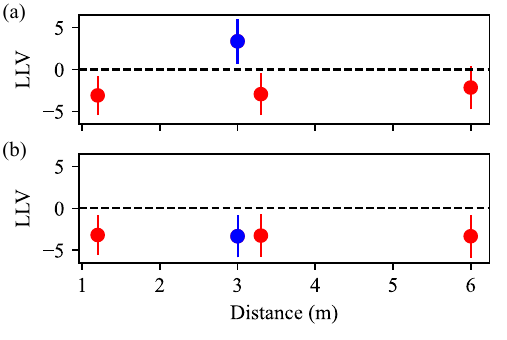}
\caption{LLV of simulated rangefinding statistics a) when an object is situated at a distance 3~m and b) when an object is absent. Mean and one standard deviation error bar plus and minus the mean plotted. The correct distance 3~m is shown as a blue dot and error bar. The shots required for threshold distinguishability $N_{\text{t}}$, signal stream delay in shots $M_{\text{delay}}(D)$ for each object distance, and $\xi$ is given in Tab.~\ref{tab:table1}. The horizontal (black) dashed line is when the LLV=0 $\bar{n}=2.19\times10^{-2}$, $\eta_\mathrm{S/I}=0.5$, signal attenuation is modelled by the reflection of a perfect Lambertian scatterer at distance D,  $\bar{n}_{\text{B,S}}=5.06\times10^{-2}$, and $\bar{n}_{\text{B,I}}=4.49\times10^{-4}$. Results of $10^4$ simulations. \label{expvsreal}}
    \end{center}\end{figure}

In the parameter regime set, for all inspected distances, the expected object present mean LLV is $\mu_{\text{E}}\approx 3.36$, knowledge of this value helps consideration towards which searched distance is correct. Simulated incoming data is processed into the LLV for that inspected distance $\Lambda_D(x,k)$ with the corresponding delay $M_{\text{delay}}(D)$. Figure~\ref{expvsreal} displays the mean LLV for each inspected distance distribution $\mu_{\text{G}}$, for both cases of an object absent and a stationary object set to be at a distance of 3~m. An error bar of one standard deviation is also plotted. These statistics have been generated from many Monte-Carlo simulation runs and each simulation run uses the same seed click data to process all inspected distances. Although it is important to note that this is not what would be seen in a real-time rangefinding protocol, as the statistics for the far distance take longer to acquire than the near distance, therefore some near distance statistics have been discarded in order to match the quantity of simulation results.
In Fig.~\ref{expvsreal} for the object present case, the correctly inspected distance 3~m shows a strong signature, due to the correct coincidence matching. However, all of the falsely inspected distances display varying levels of a shifted LLV from H0, this is due to the mismatch caused by the incoming data not adhering to \richard{the H1 or H0 statistics for that inspected distance}. Whereas, for an object absent, all \richard{inspected distances statistics match their respective object absent statistical outcome.}
    
The results from Fig.~\ref{expvsreal} can be interpreted as follows. The closer $\mu_{\text{G}}-\mu_{\text{E}}\to 0$, the closer that particular inspected distance infers that an object is actually situated at that distance. This interpretation comes with the caveat that a nearer and correct distance has not been overlooked,  as a sufficiently far searched distance can also tend $\mu_{\text{G}}-\mu_{\text{E}}\to 0$ due to the LLV shift. \richard{The searched distances are relatively near for LIDAR standards, this is because we are using an uncooperative Lambertian target in the simulation, a cooperative specular target would allow for ranging to perform well at considerably longer distances.}

\subsection{Simulated example of a rangefinding detection scenario}
\label{rolling_sample_sect}
Now that the LLV dynamics for different inspected distances have been shown, a real-time rangefinding protocol is described. 
\richard{In order to compare LLV measurements fairly from different inspected distances we must ensure that each measurement is formed from its corresponding shots required for threshold distinguishability $N_t(D)$}. Hence, for each inspected distance $D$ a LLV sample is taken every $N_{\text{t}}(D)$ shots. \richard{We define a set of $K$ LLV measurements as $\{\Lambda_1,\dots, \Lambda_K\}$, with each sample separated temporally by $N_{\text{t}}(D)$ shots}. \richard{After accumulating a number $S$ of LLV samples it is possible to calculate the sample mean, $\mu_{\text{S}}=\frac{\Sigma_{i=1}^S\Lambda_i}{S}$}.

\begin{figure}
\begin{center}
\includegraphics{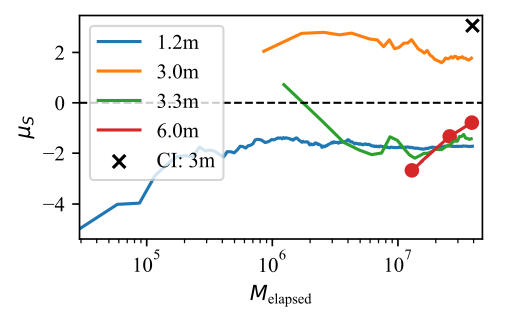}
\caption{Real-time signal trajectory of $\mu_{\text{S}}$ for the near (1.2~m), correct (3~m), one delay bin after correct (3.3~m) and far (6~m) inspected distances. With $\mu_{\text{S}}$ on the y-axis and elapsed time in shots $M_{\text{elapsed}}$ on the x-axis. The horizontal (black) dashed line is when the LLV=0. The total elapsed time in shots is $N_{\text{t}}(\text{CI:\;3~m})$. The object situated at a distance 3~m. $N_{\text{t}}$, $M_{\text{delay}}(D)$, and $\xi$ given in Tab.~\ref{tab:table1}. $f_\mathrm{rep}=0.5$~GHz, $\bar{n}=2.19\times10^{-2}$, $\eta_\mathrm{S/I}=0.5$, signal attenuation is modelled by the reflection off a perfect Lambertian scatterer at distance D,  $\bar{n}_{\text{B,S}}=5.06\times10^{-2}$, and $\bar{n}_{\text{B,I}}=4.49\times10^{-4}$. \label{realtime}}

    \end{center}\end{figure}

Plotted in Fig.~\ref{realtime} is the mean LLV $\mu_{\text{S}}(D)$ of a simulated signal for an accumulated number of $S$ LLV samples every $N_{\text{t}}(D)$ shots. \richard{The number of samples that comprise $\mu_{\text{S}}(D)$ increases with the elapsed time $M_{\text{elapsed}}$ in shots}. A $\text{3~m}$ inspected distance stream for CI is also plotted, QI's advantage is clear as the first LLV sample occurs well before CI's first sample, \richard{this means that detection decisions can be made sooner for QI than CI, which corresponds to a more responsive object detection system}. \richard{A false decision is made for the near inspected distance (1.2~m) around $M_{\text{elapsed}}\approx 5\times 10^{4}$ shots; however, it can also be seen that this inspected distance later converges to the correct decision with $\mu_{\text{G}}(\text{1.2~m})=-3.08$ relatively quickly}. Consequently, confident detection decisions are made sooner for \richard{distances searched nearest to the detector, as there is a larger number of samples for convergence to occur}. The search for an object is executed by scanning from near to far. The rangefinding protocol stops scanning outwards once there is a signature of an object at a distance $D$. Important to note, is that samples can be made sooner for any inspected distance, at the expense of detection decision error and comparability of LLV between inspected distances. \richard{Recording a sample sooner would mean that the shots $N<N_t(D)$, this would cause $P_\mathrm{D}$ to decrease and $P_\mathrm{FA}$ to increase due to the larger overlap with a smaller integration time}.

\begin{figure}
\begin{center}
\includegraphics{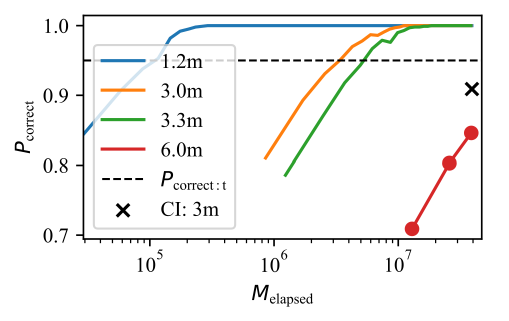}
\caption{Real-time trajectory of $\mu_{\text{S}}$ for the near (1.2~m), correct (3~m), one delay bin after correct 3.3~m and far (6~m) inspected distance. With $\mu_{\text{S}}$ on the y-axis and elapsed time $M_{\text{elapsed}}$ on the x-axis. The total elapsed time in shots is $N_{\text{t}}(\text{CI:\;3~m})$. The object situated at a distance 3~m. $N_{\text{t}}$, $M_{\text{delay}}(D)$, and $\xi$ given in Tab.~\ref{tab:table1}. $f_\mathrm{rep}=0.5$~GHz, $\bar{n}=2.19\times10^{-2}$, $\eta_\mathrm{S/I}=0.5$, signal attenuation is modelled by the reflection off a perfect Lambertian scatterer at distance D,  $\bar{n}_{\text{B,S}}=5.06\times10^{-2}$, and $\bar{n}_{\text{B,I}}=4.492\times10^{-4}$. This figure displays the statistics of $10^3$ different simulation runs, in order to generate the $\text{P}_{\text{correct}}$ distribution.\label{probcorrect}}
    \end{center}\end{figure}

An object detection decision is made \richard{when the nearest inspected distance} any $\mu_{s}>0$. Conversely, if $\mu_{s}\leq 0$ a decision is made that an object is not present at that distance. Many simulation runs of an incoming signal are performed to generate a distribution of detection decisions along elapsed time $M_{\text{elapsed}}$ in shots. For the inspected distances, excluding the furthermost, as time progresses the probability of a correct decision $P_{\text{correct}}$ improves. This is displayed in Fig.~\ref{probcorrect}. Moreover, Fig.~\ref{probcorrect} shows the number of samples required $S$ for $\text{P}_{\text{correct}}$ to reach an acceptable threshold $\text{P}_{\text{correct:t}}=0.95$. As follows, the realistic temporal resolution for a distance D when the target object is at a distance $\text{D}_{\text{correct}}$ is 
\begin{equation}
    t_{\text{realistic}}(\text{D};\text{D}_{\text{correct}})=S\times t_{\text{optimal}}(\text{D}).
\end{equation}
This trend of an improvement of $\text{P}_{\text{correct}}$ with time fails for inspected distances too far from the correct location. As the mean of a sufficiently far searched distance $D$ $\mu_{\text{G}}(D)\geq 0$. However, this situation is avoided as the correct inspected distance is decided upon before far away searches enough reach its realistic temporal resolution.

Monitoring a set of inspected distances of mean LLV's for an accumulated sample number of $S$ samples $\mu_{\text{S}}$, with a rolling window refresh rate of $S$ allows the operator to exclude the presence of an object at near distances while searching outwards to hone in on its true location. The rolling window allows for dynamic monitoring of the range of the object. Hence, the sample rolling window is defined as 
\begin{equation}
    \text{R}(\tilde{s})=\frac{1}{S}\sum_{i=\tilde{s}-S+1}^{\tilde{s}}\Lambda_i, \label{rolling_sample}
\end{equation} 
for any sample number $\tilde{s}$ such that $S\leq\tilde{s}\leq K$. This moving average rolling window defined in Eq.~\ref{rolling_sample} is what is practically used when analysing real data, as signal processing with the rolling window defined by shots in Eq.~\ref{rolling_shots} is unfeasible due to the very short temporal duration of a shot. An incoming signal is more easily processed with the sample rolling window applied. This leads to the definition of the average distinguishability, which is the distinguishability of object present and absent LLV distributions with a moving average of $S$ samples applied, 
\begin{eqnarray}
    \phi_{\text{average}}(S)&=&1-\bigl( (1-\Phi(0,\mu_{\text{H1}:\Lambda(x,\mu_\mathrm{I})},\frac{\sigma_{\text{H1}:\Lambda(x,\mu_\mathrm{I})}}{\sqrt{S}})+\nonumber \\ &+&\Phi(0,\mu_{\text{H0}:\Lambda(x,\mu_\mathrm{I})},\frac{\sigma_{\text{H0}:\Lambda(x,\mu_\mathrm{I})}}{\sqrt{S}})\bigr).
\end{eqnarray} 
Clearly when $S=1$ this equation reduces to the distinguishability defined in Eq.~\ref{single-shot_distinguishability} after mean idler clicks.

\section{Discussion}
In this paper the theory for target object detection and rangefinding using pulsed quantum states with Geiger-mode click detectors has been provided. The detection protocol was presented in terms of quantum illumination. \richard{Comparison of quantum and classical illumination (with thermal statistics) demonstrate which system parameters ensure a quantum advantage in a particular object detection scenario.}

\richard{Our model of the source state of light and detectors was written in terms of the Fock basis formalism. Detector click probabilities were calculated from this formalism}. After many shots, from these click probabilities, click-count probability distributions are generated. These click-count distributions were used to form the log-likelihood value. This quantity provides information about the relative sizes of the probabilities that the target is present and absent, so is more meaningful for inference of a target object’s presence than the click-count by itself. Each click-count has a corresponding log-likelihood value, therefore the click-count distributions are transformed into log-likelihood value distributions. Ease of, and confidence in, detecting an object is determined by how distinguished the object present and absent log-likelihood distributions are. This log-likelihood value framework allows for multi-channel detector click data to be transformed into a single value. Therefore, this approach includes \richard{non-coincidence clicks} which is often-overlooked by other quantum illumination object detection protocols. Inference using this metric also facilitates different inspected distances of the system parameters to be compared fairly for effectiveness of the log-likelihood value test, comparison of different inspected distances is required for the rangefinding aspect of the protocol. Consideration for fair comparison is particularly desired \richard{for a Lambertian reflecting target}, as system parameters differ for each inspected distance. \richard{Moreover, this theory is also frequency-independent, allowing for extension beyond LIDAR in the optical frequency regime.}

Simulation results using the protocol reinforce the knowledge that quantum illumination with click detectors performs better than classical illumination in low signal strength and high background noise regimes. This quantum advantage persists even in lossy and noisy situations with simple click detectors capable of operating in free-space. The origin of the quantum advantage is in the photon-number correlations between the two spatially separated beams of the source state. Hence, any click measurement at the idler detector conditions the probability of a click event at the signal detector to increase. In this QI-based protocol the driving force of the quantum advantage is the incidence of coincidence clicks. These occur when both the idler and the signal detectors record a click. \richard{The idler and signal detector provides a binary data stream each of a click or a no click event. Photons from our light source will have a delay in the signal stream compared to the idler stream, whereby this delay is according to a possible target object distance. A chosen delay is set and coincidence (and non-coincidence) click counts are accumulated from these detector data streams, to which they are processed into log-likelihood values. Searching through different delays and processing detector data into log-likelihood values is how we perform rangefinding of a target object.} The correctly inspected distance, and hence delay, will return statistics that are distinguishable from the false inspected distances. This is due to the reflected light from the object that causes a signal click being properly matched with its counterpart idler click, thus recovering the non-classical photon number correlations between the two beams. 

This experimentally-motivated theoretical framework has recently been applied to demonstrate the functionality of a quantum-enhanced LIDAR protocol for performing rangefinding robust to classical jamming \cite{mateusz23}. 

Further work will involve extension to multimode quantum illumination, such as modes based on spectral information. This will improve quantum illumination’s performance. Work will also be undertaken to recognise and remove jamming attempts, which will improve the versatility of this protocol in dynamic and perhaps hostile situations.

\label{sect6}

\section*{Acknowledgements}
The authors thank Mateusz P. Mrozowski for useful discussions. This work is funded by the UK Ministry of Defence. The data presented in this work are available at \cite{murchie23data}. 

\clearpage
\appendix
\section{TMSV derivation}
Derivation of the TMSV state density operator $\hat{\rho}_{\text{TMSV}}$ is presented here. Defining a complex number $\zeta=\alpha_\mathrm{P} \chi^{(2)}$, where the pump field amplitude is $\alpha_\mathrm{P}$ and the second order nonlinearity coefficient of the medium which causes the nonlinear optical process is $\chi^{(2)}$ \cite{Mollow1967}. 
Defining $\hat{K}_{+}=\hat{a}^\dag\hat{b}^\dag$, $\hat{K}_{-}=\hat{K}^\dag$ and $\hat{K}_3=\frac{1}{2}(\hat{a}^\dag \hat{a}+\hat{b} \hat{b}^\dag)$, allows the squeezing operator to be expressed in an initial form and then recast into an ordered operator form for ease of calculation with $\zeta=r e^{i\theta}$ in polar form \cite{Barnett2002} 
\begin{align}
\hat{S}(\zeta)&=e^{-\zeta \hat{K}_{+}+\zeta^* \hat{K}_{-}},\nonumber \\
&=e^{-e^{i\theta}\text{tanh}(r)\hat{K}_{+}}e^{\text{ln}(\text{cosh}^{-2}(r))\hat{K}_3}\times^{e^{-i\theta}\text{tanh}(r)\hat{K}_-}.
\end{align}
The TMSV wave-vector is hence found by applying the squeezing operator to the two-mode vacuum state, where the mode labelling $a$ and $b$ follows the ladder operators defined earlier $\vert \psi \rangle_{\text{TMSV}}= \hat{S}(\zeta)\vert 0\rangle_a \otimes \vert 0 \rangle_b\equiv\hat{S}(\zeta)\vert \{0\}_2 \rangle.$
Piece-wise deriving the following terms 

\begin{align}
&e^{e^{-i\theta}\text{tanh}(r)\hat{K}_{-}}\vert \{0\}\rangle\nonumber\\ &\quad=\sum^\infty_{n=0}\frac{(e^{-i\theta}\text{tanh}(r))^n (\hat{b}\hat{a})^n}{n!}\times\vert \{0\}_2\rangle=\vert \{ 0 \}_2 \rangle.
\end{align}

The only non-zero term is when $n=0$ and hence the vacuum is recovered. 
\begin{eqnarray}
    e^{\text{ln}(\text{cosh}^{-2}(r))\hat{K}_3}\vert \{ 0\}_2\rangle &=& \sum^\infty_{n=0}\frac{\bigl(\frac{1}{2}\text{ln}(\text{cosh}^{-2}(r))\bigr)^n}{n!}\nonumber \\ &\times& \sum^n_{q=0} {{n}\choose{q}}(\hat{b}\hat{b}^\dag)^{n-q}(\hat{a}^\dag\hat{a})^q \nonumber \\ &\times&\vert \{0\}_2\rangle, \\ &=&\sum^\infty_{n=0}\frac{(\text{ln}(\text{cosh}^{-1}(r)))^n}{n!}\nonumber \\ &\times&\vert \{ 0 \}_2 \rangle, \\ &=&\text{Sech}(r)\vert \{ 0 \}_2 \rangle.
\end{eqnarray} 
As the only non-zero term is when $q=0$. Lastly, for the piece-wise derivations 
\begin{equation}
    e^{-e^{i\theta}\text{tanh}(r)\hat{K}_{+}}\vert \{0 \}_2\rangle = \sum^\infty_{n=0}(-e^{i\theta}\text{tanh}(r))^n \vert n, n \rangle.
\end{equation}
Therefore, the TMSV wavevector 
\begin{eqnarray}
    \vert \psi \rangle_{\text{TMSV}}&=& \text{sech}(r)\sum^\infty_{n=0}(-e^{i\theta}\text{tanh}(r))^n\vert n,n\rangle, \\ &=& \frac{1}{\sqrt{\bar{n}+1}}\sum^\infty_{n=0}\left(-e^{i\theta}\sqrt{\frac{\bar{n}}{\bar{n}+1}}\right)^n \nonumber \\ &\times&\vert n,n \rangle.
\end{eqnarray} 
The relation $\bar{n}=\text{sinh}^2(r)$ is used, this relates the mean photon number with the pump field amplitude and nonlinearity coefficient of the medium. Therefore, the density matrix for the state used in QI is 
\begin{equation}
    \hat{\rho}_{\text{TMSV}}= \sum^\infty_{n=0}\frac{\bar{n}^n}{(\bar{n}+1)^{n+1}}\vert n \rangle_{\text{S}}\langle n \vert \otimes \vert n \rangle_{\text{I}}\langle n \vert.
\end{equation} 
Conventionally, the mode labelling $\text{S}$ signifies the signal mode and $\text{I}$ signifies the idler mode.
\section{POVM derivation}
\label{beamsplitter}
Deriving the generic no-click POVM element for a given mode. The parameter $\zeta$ is used to represent a generic factor which accounts for all factors of loss of the signal state. Additionally, a generic background noise mean photon number $\bar{n}_{\text{B}}$ is used here. Here, the mode labelling is mode $0$ for the background noise port, mode $1$ for the light source state port, mode $2$ for the neglected port, and mode $3$ for the click detector port. The probability of a no-click event is given by
\begin{equation}
    \text{Pr}_{\text{x}}=\text{Tr}\left(\text{Tr}_2(\hat{U}\hat{\rho}_\text{th}\otimes \hat{\rho} \hat{U}^\dag)\vert 0 \rangle_3\langle 0 \vert \right),
    \label{A1eq1}
\end{equation} 
where the beamsplitter operator is defined as \begin{equation}
   \hat{U}=e^{i\;\text{arccos}(\sqrt{\zeta})(\hat{a}^\dag_0\hat{a}_1+\hat{a}_0\hat{a}^\dag_1)}.
\end{equation}
The probability of no-click can also be defined by using a POVM element
\begin{equation}
    \text{Pr}_{\text{x}}=\text{Tr}\left(\hat{\rho}\hat{\pi}_{\times,3}\right).
\end{equation} Rearranging~\ref{A1eq1} to find the no-click POVM element for mode $3$\begin{equation}
    \hat{\pi}_{\times,3}=\text{Tr}_0\left( (\hat{\rho}_{\text{th}}\otimes \hat{1}) \hat{U}^\dag( \hat{1}\otimes\vert 0 \rangle\langle 0 \vert  )\hat{U} \right). \label{A1eq3}
\end{equation} The following derivation finds a closed-form summation over the Fock basis for Eq.~\ref{A1eq3}. The use of the coherent state basis throughout this derivation will be of great use. Firstly, 
\begin{eqnarray}
    \hat{U}^\dag (\hat{1}\otimes \vert 0 \rangle\langle 0 \vert)\hat{U}&=\frac{1}{\pi}\hat{U}^\dag \int d^2 \alpha \vert \alpha \rangle_2\langle \alpha \vert \otimes \vert 0 \rangle_3\langle 0 \vert \hat{U}. \nonumber \\ =\frac{1}{\pi}\int d^2 &\alpha \vert t^*\alpha\rangle_0\langle t^*\alpha\vert \otimes \vert r^*\alpha\rangle_1\langle r^*\alpha \vert,\label{bs1}\end{eqnarray}
    and 
    \begin{equation}
    \hat{\rho}_{\text{th}}\otimes \hat{1} = \frac{1}{\pi \bar{n}_{\text{B}}}\int d^2 \beta \vert \beta \rangle_0\langle \beta \vert \otimes \hat{1}_1.\label{bs2}
\end{equation}
Hence, with substitution of Eq.~\ref{bs1} and Eq.~\ref{bs2} into Eq.~\ref{A1eq3} yields
\begin{eqnarray}
    \hat{\pi}_{\times,3} &=&\frac{1}{\pi^3\bar{n}_{\text{B}}}\int d^2\beta d^2\alpha d^2 \gamma e^{\frac{-\vert \beta \vert^2}{\bar{n}_{\text{B}}}}\nonumber \\ 
    &\times& \langle \gamma \vert \beta \rangle\langle\beta \vert t^*\alpha\rangle\langle t^*\alpha \vert \gamma \rangle\otimes \vert r^*\alpha \rangle_3\langle r^*\alpha \vert,  \\ &=&\frac{1}{\pi^2 \bar{n}_{\text{B}} \vert r \vert^2}\int  e^{-\vert \beta \vert^2 \frac{\bar{n}_{\text{B}}+1}{\bar{n}_{\text{B}}}-\frac{\vert t\vert^2}{\vert r \vert^2} \vert \tilde{\alpha}\vert^2+\beta^* \frac{t^*}{r^*} \tilde{\alpha}}\nonumber \\ &\times& e^{\beta \frac{t}{r}\tilde{\alpha}^*}d^2 \tilde{\alpha} d^2 \beta \vert \tilde{\alpha}\rangle_3\langle\tilde{\alpha}\vert.
\end{eqnarray}
In the above derivation a change of variable is employed $\tilde{\alpha}=r^* \alpha$, hence $d^2 \alpha=\frac{d^2 \tilde{\alpha}}{\vert r \vert^2}$, and thus $\gamma=\gamma_r+i \gamma_i$ is decomposed into real and imaginary parts to compute this integral as $d^2\gamma=d\mathcal{R}(\gamma )d\mathcal{I}(\gamma)$, which allows each component to be expressed as a well known Gaussian integral for an analytic solution. Similarly, the $d^2 \beta$ integral is solved by  this real and complex decomposition.
\begin{equation}
\hat{\pi}_{\times,3}=\frac{1}{\pi \vert r \vert^2 (\bar{n}_{\text{B}}+1)}\int d^2 \tilde{\alpha} e^{-\vert \tilde{\alpha}\vert^2 \frac{\vert t\vert^2}{\vert r\vert ^2}\frac{1}{\bar{n}_{\text{B}}+1}}  \vert \tilde{\alpha} \rangle_3\langle \tilde{\alpha}\vert. \label{pre_bbar}
\end{equation} 
Setting $\bar{b}=\frac{\vert r \vert^2}{\vert t \vert^2}(\bar{n}_{\text{B}}+1)$ displays Eq.~\ref{pre_bbar} in the P-representation of a single-mode thermal state with mean photon number $\bar{b}$ with a factor of $\vert t \vert^{-2}$ in front \cite{Sudarshan1963}
\begin{equation}
    \hat{\pi}_{\times,3}=\frac{1}{\pi \vert t \vert^2 \bar{b}}\int d^2 \tilde{\alpha} e^{\frac{-\vert \tilde{\alpha}\vert^2}{\bar{b}}} \vert \tilde{\alpha}\rangle\langle \tilde{\alpha}\vert. \label{Prep}
\end{equation} 
As the measured background noise thermal state mean photon number $\bar{n}_{\text{B}}^{'}$ is to be unaffected by the reflection parameter of the beamsplitter the scaled background noise thermal state mean photon number is thus defined as $\bar{n}_{\text{B}}=\frac{\bar{n}_{\text{B}}^{'}}{\vert r \vert^2}$. The transmission and reflection parameters of the beamsplitter adhere to $\vert t \vert^2+\vert r \vert^2=1$, therefore the transmission magnitude $\vert t \vert^2=\zeta$ and reflection magnitude $\vert r \vert^2=1-\zeta$. Converting Eq.~\ref{Prep} into the Fock-basis and substituting in the mean photon number scaling, the generic no-click POVM element with measured background noise $\bar{n}_{\text{B}}^{'}$ unaffected by the POVM element for mode $\text{A}$ is 
\begin{equation}
    \hat{\pi}_{\times,\text{A}}(\zeta,\bar{n}_{\text{B}}')=\frac{1}{\bar{n}_{\text{B}}'+1}\sum^\infty_{n=0}\left(\frac{\bar{n}_{\text{B}}'+1-\zeta}{\bar{n}_{\text{B}}'+1} \right)^n \vert n \rangle_{\text{A}}\langle n \vert. 
\end{equation}  Consequently, the generic click POVM element for mode $\text{A}$ is \begin{equation}
    \hat{\pi}_{\checkmark,\text{A}}(\zeta,\bar{n}_{\text{B}}')=\hat{1}_{\text{A}}-\hat{\pi}_{\times,\text{A}}(\zeta,\bar{n}_{\text{B}}').\label{generic_click_POVM}
\end{equation}
\section{Gaussian approximation} \label{gaussian} Analytic simplicity and computational speed is afforded by operating in the Gaussian regime. This regime is the system parameters such that the binomial click-count distributions can be approximated as Gaussian distributions, with negligible error produced. The following criteria is applied to ensure the Gaussian approximation is valid for the system parameters set. The criteria is that all binomial distributions for the system parameters in question must not be too skewed. In other words this inequality must be satisfied 
\begin{equation}
    \frac{1-2p}{\sqrt{Np (1-p)}}<0.3. \label{ineq}
\end{equation}
Where $p$ is the probability which underlies the binomial distribution and $N$ is the number of shots. In practice not all binomial distributions need to be checked, instead only the binomial distribution which is most prone to failing the criteria to be in the Gaussian regime are checked. In the case of low signal strength in a noisy and lossy environment, the weakest distribution is the object absent signal coincidence click-count distribution after a thresholded minimum of idler clicks. The mean and standard deviation of the idler click distribution is $\mu_{\text{I}}$ and $\sigma_{\text{I}}$ respectively. These statistical moments for the idler distribution are derived from the probability of an idler click $\text{Pr}_{\text{I}}$ and number of shots $N$. The thresholded minimum of idler clicks is $\text{I}_{\text{min}}=\text{int}(\mu_{\text{I}}-4\sigma_{\text{I}})$. Hence the weakest distribution has the form 
\begin{equation}
    P_{\text{min}}(x)=\binom{\text{I}_{\text{min}}}{x}p_{\text{min}}^x(1-p_{\text{min}})^{\text{I}_{\text{min}}-x},
\end{equation} 
where $p_{\text{min}}$ is the least likely type of signal click event in the analysed system when the object is absent. The least likely type of signal click event for the system is the coincidence click. If this distribution satisfies Eq.~\ref{ineq} then the Gaussian approximation is valid for the given system parameters.
Once it has been demonstrated that the Gaussian regime can be operated in, any binomial distribution pertaining to those system parameters is transformed from
\begin{equation}
    P(x,N,p)=\binom{N}{x}p^x(1-p)^{N-x}.
\end{equation} to 
\begin{equation}
    P(x,N,p)\approx \frac{1}{\sigma\sqrt{2\pi}}e^{-0.5(\frac{x-\mu}{\sigma})^2},
\end{equation}
where $\mu=Np$ and $\sigma=\sqrt{Np(1-p)}$.
\section{Deriving the linear form of the LLV} \label{LLVtrans_appendix}
The generic LLV in its ratio of click probability distributions form
\begin{equation}
    \Lambda(\underline{x},k)=\text{ln}\left(\frac{P_{\text{H1}}(\underline{x},k)}{P_{\text{H0}}(\underline{x},k)}\right), \label{LLVtrans}
\end{equation} 
$\underline{x}$ is the vector of the click counts by type of click event, $k$ is the relevant number of shots (CI total shots and QI number of idler clicks), and $P_{\text{H}\{1,0\}}(\underline{x},k)$ is the probability for object present or absent respectively. The probability distribution for click events is originally binomial, due to the Bernoulli trials undertaken. For the remainder of the appendices I use shorthand notation for the click probabilities $\text{Pr}_{\text{H1:CI}}\equiv p_{\text{H1}}$, $\text{Pr}_{\text{H0:CI}}\equiv p_{\text{H0}}$, $\text{Pr}_{\text{I}}\equiv p_{\text{I}}$, $\text{Pr}_{\text{H1:I,1}}\equiv p_{\text{H1:I1}}$, $\text{Pr}_{\text{H1:I,0}}\equiv p_{\text{H1:I0}}$, and $\text{Pr}_{\text{H0}}\equiv p_{\text{H0:I1}} \equiv p_{\text{H0:I0}}$. The simpler CI case is focused on first, as there is only one element in $\underline{x}\equiv x$. The object present and absent probability density functions in its binomial form, respectively 
\begin{eqnarray}
    P_{\text{H1}}(x,N)&=&{{N}\choose{x}}p_{\text{H1}}^x(1-p_{\text{H1}})^{N-x}, \\
    P_{\text{H0}}(x,N)&=&{{N}\choose{x}}p_{\text{H0}}^x(1-p_{\text{H0}})^{N-x}.
\end{eqnarray}
As $N,x$, and all the click probabilities are all real and positive Eq.~\ref{LLVtrans} can be expressed as a linear equation. Hence, for CI Eq.~\ref{LLVtrans} is
\begin{equation}
    \Lambda(x,N)=Mx+NC, \label{lineartrans}
\end{equation} 
where $M=\text{ln}\left(\frac{p_{\text{H1}}(1-p_{\text{H0}})}{p_{\text{H0}}(1-p_{\text{H1}})}\right)$ and $C=\text{ln}\left(\frac{1-p_{\text{H1}}}{1-p_{\text{H0}}}\right)$. The LLV transformation for QI easily extends to include idler not firing events in the linear equation. The signal click-count $x$ after $k$ idler click events and signal click-count $y$ after $N-k$ idler no-firing events in the QI protocol is transformed by the LLV defined as
\begin{equation}
    \Lambda(\underline{x},k)=\underline{M}^{\text{T}}\underline{x}+\underline{N}\underline{C}^{\text{T}},\label{QIvectlintrans}
\end{equation}
where $\underline{M}^{\text{T}}=( M_1 \; M_2 )$, $\underline{x}=(x \; y )$, $\underline{C}^{\text{T}}=(C_1 \; C_2)$, and $\underline{N}={k \choose N-k}$. Where 
\begin{eqnarray}M_1&=&\text{ln}\left(\frac{p_{\text{H1:I1}}(1-p_{\text{H0:I1}})}{p_{\text{H0:I1}}(1-p_{\text{H1:I1}})}\right), \\ M_2&=&\text{ln}\left(\frac{p_{\text{H1:I0}}(1-p_{\text{H0:I0}})}{p_{\text{H0:I0}}(1-p_{\text{H1:I0}})}\right), \\ C_1&=&\text{ln}\left(\frac{1-p_{\text{H1:I1}}}{1-p_{\text{H0:I1}}}\right), \\  C_2&=&\text{ln}\left(\frac{1-p_{\text{H1:I0}}}{1-p_{\text{H0:I0}}}\right).
\end{eqnarray}.

\section{Click to LLV distribution}
The condition that the Gaussian regime is valid ensures that all click-count distributions can easily be transformed into its LLV form. This ease is afforded by the LLV for CI Eq.~\ref{lineartrans} and QI Eq.~\ref{QIvectlintrans} being linear, this property preserves normality. The first two statistical moments of any click-count distribution are transformed such that the first two statistical moments are yielded for the LLV distributions. The following analysis is presented for the object present case. Deriving for the object absent case instead requires replacing the relevant click probabilities. Shown below is the transformation for the mean $\mu_{\text{H1:CI}}=N p_{\text{H1}}$ and standard deviation $\sigma_{\text{H1:CI}}=\sqrt{Np_{\text{H1}}(1-p_{\text{H1}})}$ for CI 
\begin{eqnarray}
    \mu_{\text{H1:CI:}\Lambda}&=&\Lambda(\mu_{\text{H1:CI}},N), \\
    &=&M\mu_{\text{H1:CI}}+NC.
\end{eqnarray} and 
\begin{eqnarray}
    \sigma_{\text{H1:CI:}\Lambda}&=M\sigma_{\text{H1:CI}}.
\end{eqnarray}
However, it is not as simple for QI. Equation~\ref{QIvectlintrans} encodes both coincidence clicks and non-coincidence click-count distributions into the one LLV distribution. Equation~\ref{QIvectlintrans} amounts to a linear combination, therefore normality is preserved for the resulting LLV distribution. If there has been $k$ idler click events, a click-count distribution's statistical moments $(\mu_{\text{H1:}k},\sigma_{\text{H1:}k})$ is transformed into its respective LLV statistical moments as 
\begin{eqnarray}
    \mu_{\text{H1}:\Lambda(x,k)}&=& M_1 k p_{\text{H1:I1}}+C_1 k + M_2\left((N-k)p_{\text{H1:I0}}\right)+ \nonumber \\ &+&(N-k)C_2.
\end{eqnarray} 
for the mean and 
\begin{eqnarray}
    \sigma_{\text{H1:}\Lambda(x,k)}&=& \bigl(M_1^2 k p_{\text{H1:I1}}(1-p_{\text{H1:I1}})+\nonumber \\
    &+&M_2^2(N-k)p_{\text{H1:I0}}(1-p_{\text{H1:I0}})\bigr)^{0.5}
\end{eqnarray} for the standard deviation.
Much of the characterisation of system performance is oriented around the LLV distributions after mean idler clicks $\mu_{\text{I}}=Np_{\text{I}}$, rather than for $k$ idler clicks. Therefore, the mean of the object present LLV distribution after mean idler clicks is defined as
\begin{eqnarray}
    \mu_{\text{H1:}\Lambda(x,\mu_\mathrm{I})}&=& N\bigl(p_{\text{I}}(M_1 p_{\text{H1:I1}}+C_1-M_2 p_{\text{H1:I0}}-C_2)+\nonumber\\&+&M_2 p_{\text{H1:I0}}+C_2\bigr).
\end{eqnarray}

The standard deviation of the object present LLV distribution after mean idler clicks is 
\begin{eqnarray}
    \sigma_{\text{H1:}\Lambda(x,\mu_\mathrm{I})}&=&\biggl( N\bigl(p_{\text{I}}(M_1^2 p_{\text{H1:I1}}(1-p_{\text{H1:I1}})+\nonumber \\
    &-& M_2^2 p_{\text{H1:I0}}(1-p_{\text{H1:I0}}))+\nonumber \\ &+& M_2^2p_{\text{H1:I0}}(1-p_{\text{H1:I0}})\bigr)\biggr)^{0.5}.
\end{eqnarray}

\label{click_to_LLV}
\section{FOM comparison}
\label{appendix_FOM}
The FOM's, distinguishability $\phi$, SNR, and CRLB for signal loss estimation, when compared illustrates why the distinguishability $\phi$ is suitable in the detection-decision orientated theory. Comparing performance of QI and CI as a ratio of their respective distinguishabilities in certain regimes is vulnerable to the issues of divergence ($\phi_{\text{t:CI}}=0$) or saturation ($\phi_{\text{t:QI,CI}}=1$). Instead, the ratio of the QI and CI shots required $N_\text{t}$ for a distinguishability $\phi$ is used in this paper. While the ratio of QI and CI $N_{\text{t}}$ does not avoid the issue of divergence ($\phi_{\text{t:CI}}=0 \equiv N_{\text{t:CI}}\to \infty$) in certain regimes, it frames performance as a ratio of time required for confident detection, which is intuitive. Regardless, comparing the values of the QI and CI distinguishabilities still functions as a method of comparing system performance.

The SNR can yield the same result for multiple different regimes, hence it has no uniqueness, therefore it is insufficient for characterising detection decision performance.

Moreover, the CRLB for signal attenuation estimation, while well grounded in estimation theory and developed for the context of quantum illumination photo-counting in Ref.~\cite{Liu2019}, it does not directly instruct whether a confident decision can be made or not.

This appendix focuses on QI for the FOM's, as it is straight-forward to apply these approaches to CI. The SNR is defined as \begin{equation}
    SNR=\frac{N_{\text{coincidences}}}{N_{\text{noise}}},
\end{equation}where $N_{\text{coincidences}}$ is the number of coincidence clicks and $N_{\text{noise}}$ is the number of background noise clicks.
Signal attenuation is estimated with $\xi$. The following analysis can proceed assuming that the signal attenuation estimation is greater than the uncertainty of the signal attenuation estimation. The CRLB instructs system performance by stating how easy an object's presence can be discerned, subject to noise and the variance of the underlying distribution. It is defined, for an estimator of signal attenuation $\hat{\xi}$, that the CRLB is \begin{equation}
    \Delta^2 \hat{\xi}=\mathbb{E}\biggl(\frac{-\partial^2 \text{ln} P_{\text{H1}}(\xi)}{\partial (\xi)^2} \biggr)^{-1}.
\end{equation} Following from this, the FOM is given with respect to the estimated signal attenuation and in decibels as \begin{equation}
    \Delta^2 \hat{\xi}\; (\text{dB})= 10\text{log}_{10}\left(\frac{\Delta^2\hat{\xi}}{\hat{\xi}}\right).
\end{equation}
\section{Distinguishability discrepancy}

The LLV after $k$ idler clicks processes coincidence (and possibly non-coincidence) click data which has had $k$ idler clicks, once this processing has occurred the knowledge of the underlying click data is obscured and only a LLV is known. \richard{Neglecting our knowledge of how many idler clicks there has been simplifies the post-processing of LLV data. Therefore, for any LLV after $k$ idler clicks this LLV can be processed with any other LLV with $\tilde{k}\neq k$ idler clicks. In other words, each LLV after any number of idler clicks are equivalent to each other in post-processing.} Hence, it is important the discrepancy in the effectiveness of each LLV test after $k$ idler clicks is limited. Otherwise this equivalency is erroneous. 

Given a single regime, QI has object present and absent LLV distributions for each $k$ idler clicks. This corresponds to a distinguishability for each LLV after $k$ idler clicks, which is denoted as $\phi_{k}$. The system performance is characterised in terms of the threshold distinguishability $\phi_{\text{t}}$, which is calculated from the LLV distributions after mean idler clicks. Consequently, there must be only a limited discrepancy between any $\phi_{k}$ and $\phi_{\text{t}}$, for LLV equivalency to be valid.

Bounds are placed on what is considered to be extremal numbers of $k$ idler clicks. With the minimum and maximum $k$ idler clicks being set as $\text{I}_{\text{min}}=\text{int}(\mu_{\text{I}}-4\sigma_{\text{I}})$ and $\text{I}_{\text{max}}=\text{int}(\mu_{\text{I}}+4\sigma_{\text{I}})$, respectively. With $\mu_{\text{I}}$ as the mean and $\sigma_{\text{I}}$ as the standard deviation of the idler clicks distribution. 

The condition for limited discrepancy is arbitrarily set as $\text{T}_{\phi}=0.05$. Therefore, a regime has an acceptable amount of distinguishability discrepancy if both criterion \begin{equation}\frac{\vert \phi_{\text{t}}-\phi_{\text{I}_{\text{min}}}\vert}{\phi_{\text{t}}} \leq\text{T}_{\phi} \;\; \text{and}\;\; \frac{\vert \phi_{\text{t}}-\phi_{\text{I}_{\text{max}}}\vert}{\phi_{\text{t}}}\leq \text{T}_{\phi}\end{equation} are satisfied.
\label{appendix_LLV_k_consistency}
\section{Solving distinguishability equation}
The distinguishability measure for the LLV distributions after mean idler clicks is defined as 
\begin{eqnarray}
    \phi&=&1-\bigl( (1-\text{Q}(0,\mu_{\text{H1}:\Lambda(x,\mu_\mathrm{I})},\sigma_{\text{H1}:\Lambda(x,\mu_\mathrm{I})}))+\nonumber \\
    &+&\text{Q}(0,\mu_{\text{H0}:\Lambda(x,\mu_\mathrm{I})},\sigma_{\text{H0}:\Lambda(x,\mu_\mathrm{I})})\bigr).
\end{eqnarray}
where $\text{Q}(d_{\text{LLV}},\mu,\sigma)$ is the Gaussian Q-function. The statistical moments $\mu_{\Lambda(x,\mu_\mathrm{I})}$ and $\sigma_{\Lambda(x,\mu_\mathrm{I})}$ are from the LLV distributions for object present or absent.
The Gaussian Q-function can be approximated by the error function $\text{Erf}(d_{\text{LLV}})$ as 
\begin{equation}
    \text{Q}(d_{\text{LLV}},\mu,\sigma)=0.5\left(1-\text{Erf}(\frac{d_{\text{LLV}}-\mu}{\sigma\sqrt{2}})\right).
 \end{equation} 
 Hence, the definition of $\phi$ is restated as
\begin{equation}
    \phi=0.5\left(\text{Erf}(\frac{-\mu_{\text{H0}:\Lambda}}{\sigma_{\text{H0}:\Lambda}\sqrt{2}})+\text{Erf}(\frac{\mu_{\text{H1}:\Lambda}}{\sigma_{\text{H1}:\Lambda}\sqrt{2}})\right). \label{distinguish_eqn}
\end{equation} 
The signage for above is dictated by the need for a positive argument in the error function. Solving Eq.~\ref{distinguish_eqn} to find the parameters required for $\phi=\phi_{\text{t}}$ needs the shots required for threshold distinguishability $N_{\text{t}}$ to be found. Firstly, for QI the shots required is decomposed into $N_{\text{t}}=N_{\text{I1}}+N_{\text{I0}}$ and $N_{\text{t}}=\frac{N_{\text{I1}}}{p_{\text{I}}}$, where $N_{\text{I1}}$ and $N_{\text{I0}}$ are shots when the idler does and does not fire, respectively. From the prior definitions it is clear that $N_{\text{I0}}=N_{\text{I1}}(\frac{1}{p_{\text{I}}}-1)$. Eq.~\ref{distinguish_eqn} framed in terms of the variable $N_{\text{I1}}$ when $\phi=\phi_{\text{t}}$ is expressed as 
\begin{equation}
\phi_{\text{t}}=0.5\left(\text{Erf}(-G_0\sqrt{N_{\text{I1}}})+\text{Erf}(G_1\sqrt{N_{\text{I1}}})\right), \label{distinguish_eqn2}
\end{equation} where 
\begin{equation}
G=\frac{M_1 p_{\text{I1}}+C_1+(\frac{1}{p_{\text{I}}}-1)(M_2 p_{\text{I0}}+C_{2})}{\sqrt{2}\left(M_1^2 p_{\text{I1}}(1-p_{\text{I1}})+M_2^2(\frac{1}{p_{\text{I}}}-1)p_{\text{I0}}(1-p_{\text{I0}})\right)^{0.5}}.\nonumber
\end{equation} Replacing object present and absent click probabilities to yield $G_1$ and $G_0$ respectively. Therefore, $N_{\text{I1}}$ is numerically solved with the inverse function of $\phi_{\text{t}}$
\begin{equation}
    N_{\text{I1}}=\text{F}^{-1}(\phi_{\text{t}}).
\end{equation} 
Following this, the shots required for threshold distinguishability is 
\begin{equation}
N_{\text{t}}=\text{int}\left(\frac{N_{\text{I1}}}{p_{\text{I}}}\right). 
\end{equation}

\label{solvedisting}
\bibliographystyle{apsrev4-2}

\end{document}